\documentclass[twocolumn]{aa}
\usepackage{graphicx}
\usepackage{lscape}
\usepackage{graphics}	
\usepackage{epsfig}  
\usepackage{natbib}
\bibpunct{(}{)}{;}{a}{}{,} 
\usepackage{float}
\usepackage{geometry}	
\usepackage{subfig}

\begin{document}
   \title{High-resolution observations of two O\,{\sc vi} absorbers at 
          $z\approx2$ towards PKS\,1448$-$232}

 \subtitle{}

  \author{N. Draganova,
          \inst{1}
          P. Richter,
          \inst{1}
          \and
           C. Fechner
	\inst{1}
         }

  \offprints{N. Draganova
  \email{nadia@astro.physik.uni-potsdam.de}}

   \institute{
              Universit\"at Potsdam,
              Institut f\"ur Physik und Astronomie,
              Karl-Liebknecht-Strasse 24-25, Haus 28,
              14476 Potsdam, Germany\\
              \email{nadia@astro.physik.uni-potsdam.de}
              }

   \date{Received xxx, 2011; accepted xxx}


   \abstract{
To explore the ionization conditions in highly-ionized absorbers 
at high redshift we have studied in detail two intervening 
O\,{\sc vi} absorbers at $z\approx 2$ towards the quasar PKS\,1448$-$232, 
based on high ($R\approx 75,000$) and intermediate ($R\approx 45,000$) 
resolution optical VLT/UVES spectra. We find that both absorption
systems are composed of several narrow subcomponents with C\,{\sc iv}/O\,{\sc vi}
Doppler-parameters $b<10$ km\,s$^{-1}$, typically. This implies
that the gas temperatures are $T<10^5$ K and that the 
absorbers are photoionized by the UV background.
The system at $z=2.1098$ represents a simple, isolated 
O\,{\sc vi} absorber that has only two absorption components and that 
is relatively metal-rich ($Z\sim 0.6$ solar). 
Ioinization modeling implies that the
system is photoionized with O\,{\sc vi}, C\,{\sc iv}, and H\,{\sc i} coexisting in
the same gas phase. 
The second system at $z=2.1660$ represents a complicated,
multi-component absorption system with eight O\,{\sc vi} components spanning almost
$300$ km\,s$^{-1}$ in radial velocity. The photoionization modeling implies 
that the metallicity is non-uniform and relatively low ($\leq0.1$ solar) and 
that the O\,{\sc vi} absorption must arise in a gas phase different from that
traced by C\,{\sc iv}, C\,{\sc iii}, and H\,{\sc i}. 
Our detailed study of the two O\,{\sc vi} systems towards PKS\,1448$-$232
shows that multi-phase, multi-component high-ion absorbers like the 
one at $z=2.1660$ require a detailed ionization modeling of the various 
subcomponents to obtain reliable results on the physical conditions
and metal-abundances in the gas. 

   }
   \titlerunning{O\,{\sc vi} systems towards PKS\,1448$-$232}
   \maketitle


\section{Introduction}

Highly ionized species like O\,{\sc vi} and C\,{\sc iv}, observed 
in the spectra of distant quasars, are excellent tracers of 
metal-enriched ionized gas in the filamentary intergalactic medium 
(IGM) and in the circumgalactic environment of galaxies.
Therefore, the analysis of intervening O\,{\sc vi} and C\,{\sc iv} 
absorbers towards low- and high-redshift QSOs is crucial for a 
better understanding of the physical nature, distribution, evolution, 
and baryon and metal content of the IGM in the context of 
galaxy evolution. 
Because of the high cosmic abundance
of oxygen, the large oscillator strength of the 
O\,{\sc vi} doublet (located in far-ultraviolet
at $\lambda\lambda 1031.9,1037.6$ \AA), and the
high ionization energies of the ionization
states O$^{+4}$ (113.9 eV) and O$^{+5}$ (138.1 eV),
the O\,{\sc vi} ion is a particularly powerful tracer of the
metal-enriched IGM and the gaseous environment of 
galaxies. Using QSO absorption spectroscopy, 
O\,{\sc vi} absorption now is commonly detected in
various different galactic and intergalactic 
environments in the redshift range 
$z\approx 0-3$.

In the local Universe, O\,{\sc vi} absorption in interstellar and
intergalactic gas can be observed in the
FUV spectra of stars and extragalactic background sources. For
instance, O\,{\sc vi} absorption is known to arise
in the thick disk of the Milky Way \citep[e.g.][]{Savage03}, 
in the extended, multi-phase gas halos of the Milky Way and other
galaxies \citep[e.g.][]{Sembach03, Wakker09, Prochaska11}, 
and in intervening O\,{\sc vi} absorption-line systems that trace 
metal-enriched gas in the IGM 
\citep[e.g.,][]{Tripp, Savage02, Richter04, Sembach04,
Tripp08, Thom08a, Thom08b, Danforth08}, 
for a review see \citet{Richter08}.
Over the last decade, intervening O\,{\sc vi} absorbers 
at low redshift were considered as major
baryon reservoir in the IGM, possibly tracing
shock-heated and collisionally ionized intergalactic gas that 
results from large-scale structure formation \citep{Cen99, Dave}.
This so-called warm-hot intergalactic medium (WHIM) has gas
temperatures in the range $10^{5}<T<10^{7}$ K and
is believed to host $30-40\%$ of the baryons at $z=0$ \citep{Cen99}.
Recent observational and theoretical studies indicate, however, that
part of the O\,{\sc vi} absorbers at low $z$ may trace low-density, 
photoionized gas or conductive, turbulent, or shocked boundary layers
between cold/warm ($\sim 10^3-10^4$ K) gas clouds and an ambient
hot ($\sim 10^6-10^7$ K) plasma rather than the shock-heated WHIM 
\citep[see discussion in]{Fox11}. Thus, a simple estimate
of the ionization state of the gas in the absorbers 
from the observed O\,{\sc vi}/H\,{\sc i} ratios may lead 
to erroneous results because of the complex
multi-phase character of the gas \citep{Tepper-Garcia11}.

For redshifts $z>2$ O\,{\sc vi} absorption is detectable
from the ground, where it can be observed in optical QSO
absorption spectra at relatively high signal-to-noise (S/N). 
One very problematic aspect for the analysis of 
O\,{\sc vi} absorbers at high redshift is
the often severe blending of the O\,{\sc vi} absorption
with the Ly\,$\alpha$ forest. As for low redshifts, the origin and
nature of O\,{\sc vi} absorbers at high $z$ is expected to be 
manifold. It has been shown by simulations \citep[e.g.][]{Theuns02, 
Oppenheimer08} that shock-heating by collapsing large-scale structures 
is not efficient at high redshift to provide a widespread warm-hot intergalactic 
phase in the early Universe. Instead, galactic winds probably 
contribute substantially to the population of photoionized and 
collisionally ionized O\,{\sc vi} absorbers at high redshifts, 
enriching the surrounding circumgalactic and intergalactic gas
with heavy elements at relatively high gas temperatures 
\citep {Fangano07, Kawata07}. In fact, many of the strong
O\,{\sc vi} absorbers at high $z$ exhibit complex absorption
patterns that would be expected for a circumgalactic multi-phase
gas environment \citep[e.g.][]{Bergeron05}. Similar as for 
circumgalactic absorbers in the local Universe, a considerable fraction 
of the O\,{\sc vi} absorbers at high $z$ thus may arise in
conductive, turbulent, or shocked boundary layers.

Next to those O\,{\sc vi} absorbers that trace highly-ionized gas
in the immediate environment of galaxies, intergalactic 
O\,{\sc vi} absorbers (i.e., absorbers that are not gravitationally
bound to individual galaxies) may arise in regions
that are sufficiently enriched with heavy elements. 
Previous surveys of high-redshift O\,{\sc vi} absorbers
\citep{Bergeron02, Simcoe02, Simcoe04, Simcoe06, 
Carswell02, Bergeron05} have shown that
there are many narrow O\,{\sc vi} absorbers with Doppler-parameters
$b$ $\leq10$ km\,s$^{-1}$. Such narrow lines cannot arise from
collisionally ionized gas but must be
related to photoionized (possibly intergalactic) gas with 
temperatures $T<10^5$ K. Many of these narrow O\,{\sc vi} 
absorbers at low and high redshift display velocity-centroid 
offsets between O\,{\sc vi}, C\,{\sc iv}, and H\,{\sc i},
suggesting that these ions do not arise in the same gas
phase. Unfortunately, this crucial aspect is only partially
considered in previous O\,{\sc vi} surveys.

To explore the multi-phase character of high-ion absorbers 
and to improve our understanding of the ionization conditions in 
O\,{\sc vi} systems it is important to investigate in detail 
the absorption characteristics and ionization conditions in 
{\it selected} absorption-line systems. For this purpose, absorbers
that can be observed at high S/N and
for which the O\,{\sc vi} absorption is not blended by Ly\,$\alpha$
forest lines are particularly important.
Because of the complexity of many high-ion absorbers that
often are composed of several velocity subcomponents a spectral
resolution of $R\approx45,000$ and higher is desired.
Note that while the analysis of individual
high-ion absorbers is a common strategy to explore the nature
of O\,{\sc vi} absorbers at low redshift 
\citep[e.g.][]{Tumlinson11, Savage11}, detailed studies of individual O\,{\sc vi} 
absorption systems at high redshift are rare \citep[e.g.,][]{Fox11a}.

In this paper we present VLT/UVES observations at intermediate
($R\approx45,000$) and high ($R\approx75,000$)
spectral resolution of two particularly interesting 
O\,{\sc vi} systems at $z\approx2$ along the line of sight 
towards the quasar PKS 1448$-$232. This sightline was selected
by us for a detailed study, as it contains two unsaturated
O\,{\sc vi} systems at $z_{\rm abs}=2.1098$ and
$z_{\rm abs}=2.1660$, both displaying a well-defined 
subcomponent structure with narrow O\,{\sc vi}/C\,{\sc iv}
absorption components and without major blending with
Ly\,$\alpha$ forest lines \citep{Bergeron02,Fox08}.
These two absorption systems therefore represent 
ideal targets to study in detail the physical conditions in
photoionized, multi-phase high-ion absorbers 
at high redshift.

\begin{figure*}[!th]
\begin{center}
\resizebox{0.55\hsize}{!}{\includegraphics{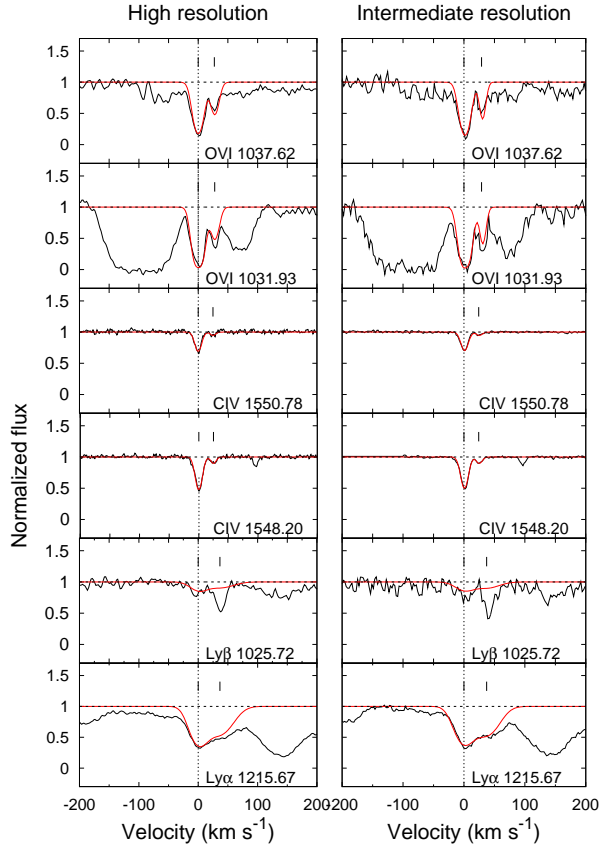}}
\caption{Absorption profiles for the O\,{\sc vi} absorber at $z=2.1098$
in the high-resolution data (left panel) and the intermediate-resolution
data (right panel).}
\end{center}
\end{figure*}


\section{Observations and absorption-line analysis}

\subsection{VLT/UVES observations}

Our data set consists of intermediate- and high-resolution 
spectra of the quasar PKS\,1448$-$232 ($z_{\rm em}=2.208$; 
$V=16.9$), observed at the VLT with the UVES spectrograph. 
The intermediate-resolution data have a spectral resolution 
of $R\approx45,000$, corresponding to a velocity resolution 
of $\Delta v \approx 6.7$ km\,s$^{-1}$ FWHM. These data were obtained and 
reduced as part of the ESO Large Programme ``The Cosmic 
Evolution of the IGM'' \citep{Bergeron02}. The wavelength 
coverage of the intermediate resolution data is $3050-10,400~\rm\AA$.
The S/N in the data varies between $15$ and $90$ per spectral 
resolution element.

The high-resolution data 
have $R\approx75,000$, corresponding to $\Delta v \approx 4.0$ km\,s$^{-1}$ 
FWHM velocity resolution and were obtained with VLT/UVES in 
2007 in an independent observing run (program ID 079.A$-$0303(A)).
The wavelength coverage of the high resolution data is $3000-6687~\rm\AA$.
The raw data were reduced using the UVES pipeline implemented in the 
ESO-MIDAS software package. The pipeline reduction includes flat-fielding, 
bias- and sky-subtraction and a relative wavelength calibration. The individual 
spectra then have been corrected to vacuum wavelengths and coadded. 
The S/N in the high-resolution data is $20-70$ per resolution element.


\begin{table*}[]
\caption{Fit parameters for the system at $z=2.1098$}
\begin{tiny}
\begin{tabular}{cccccccccc}
\hline
\hline
\\
& \multicolumn{3}{c}{\small $z$}& \multicolumn{2}{c}{\small O\,{\sc vi}} &
\multicolumn{2}{c}{\small C\,{\sc iv}} &  \multicolumn{2}{c}{\small H\,{\sc i}}\\
 & O\,{\sc vi} & C\,{\sc iv} & H\,{\sc i} & log[$N$(cm$^{-2}$)] & 
$b$\,[km\,s$^{-1}$] & log[$N$(cm$^{-2}$)] & $b$\,[km\,s$^{-1}$] &       
log[$N$(cm$^{-2}$)] & $b$\,[km\,s$^{-1}$] \\ \hline
\multicolumn{10}{c}{\it high resolution data} \\ \hline
1 & 2.10982 & 2.10982 & 2.10981 & 14.27($\pm$0.01) & 10.7($\pm$0.2) & 13.12($\pm$0.01) & 7.5($\pm$0.1) & 13.38($\pm$0.04) & 19.6($\pm$0.7) \\
2 & 2.11011 & 2.11008 & 2.11018 & 13.50($\pm$0.02) &  8.4($\pm$0.4) & 12.23($\pm$0.03) & 6.1($\pm$0.6) & 13.37($\pm$0.04) & 28.6($\pm$1.7) \\ \hline
\multicolumn{10}{c}{\it intermediate resolution data} \\ \hline
1 & 2.10984 & 2.10983 & 2.10981 & 14.32($\pm$0.02) &10.1($\pm$0.2)  & 13.12($\pm$0.01) & 7.1($\pm$0.1) & 13.39($\pm$0.01) & 20.6($\pm$0.3) \\
2 & 2.11014 & 2.11008 & 2.11019 & 13.49($\pm$0.20) & 5.4($\pm$0.5)  & 12.23($\pm$0.03) & 5.3($\pm$0.6) & 13.35($\pm$0.01) & 26.4($\pm$0.8) \\ \hline
 \end{tabular}
\end{tiny}
\end{table*}


\subsection{Line-fitting method}

The detected absorption features that are associated with the
two absorbers at $z_{\rm abs}=2.1098$ and
$z_{\rm abs}=2.1660$ were fitted independently 
in both spectra (at intermediate resolution and high resolution) 
with Gaussian profiles using the CANDALF fitting routine
\footnote{written by Robert Baade, Hamburger Sternwarte},
which uses a standard Levenberg-Marquard minimization algorithm. 
The program simultaneously fits the continuum and the absorption lines,
delivering ion column densities, $N$, and Doppler parameters, $b$,
of for each absorption component.
The continuum is modeled as a Legendre polynomial with an order of up to 4. 
The one-sigma fitting errors for $N$ and $b$ (as listed in Tables $1-3$) 
are estimated using the diagonals of the Hesse matrix. 

\begin{figure*}[!th]
\begin{center}
\resizebox{0.77\hsize}{!}{\includegraphics{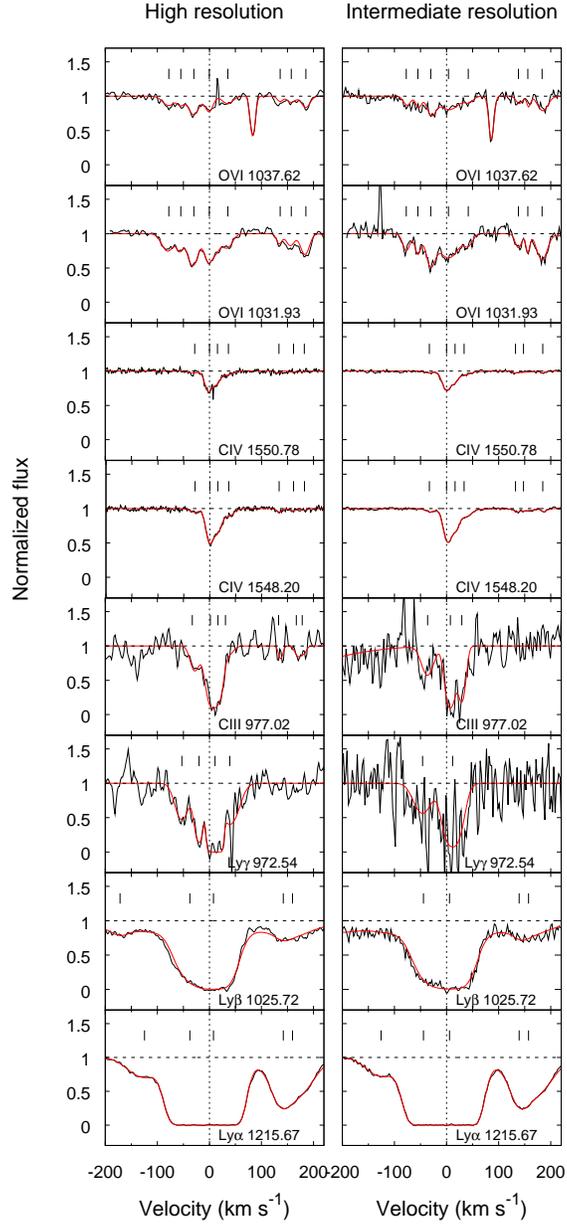}}
\caption{Absorption profiles for the O\,{\sc vi} absorber at $z=2.1660$
in the high-resolution data (left panel) and the intermediate-resolution
data (right panel). The strong absorption observed in O\,{\sc vi} $\lambda 
1037.62$ plot is a Si\,{\sc iii} line at $z=1.7236$.}
\end{center}
\end{figure*}


\subsection{The O\,{\sc vi} system at $z=2.1098$}

Fig.\,1 shows the velocity profiles of O\,{\sc vi} ($\lambda\lambda 1031.9,1037.6$),
C\,{\sc iv} ($\lambda\lambda 1548.2,1550.8$), and H\,{\sc i} Ly\,$\alpha$ and
Ly\,$\beta$ ($\lambda\lambda 1215.7,1025.7$) for the $z=2.1098$
absorber in the high-resolution data (left panel) and the intermediate-resolution
data (right panel). From the visual inspection of both panels we find
no significant differences between the two data sets. The
S/N ratios are better for the high resolution data, 
except for the C\,{\sc iv} region
where this ratio is slightly better at intermediate resolution. 
Therefore, the differences in the values for $N$, $b$, and $z$ derived for the 
individual absorption components in the intermediate and high-resolution 
spectra are a result of the different S/N values in the two data sets.

Two absorption components are detected in each of these ions. The O\,{\sc vi}
absorption is relatively strong compared to C\,{\sc iv}. H\,{\sc i}
absorption is weak compared to other O\,{\sc vi} absorbers at similar
redshift \citep[e.g.,][]{Bergeron02} with a central absorption
depth in the H\,{\sc i} Ly\,$\alpha$ line of less than 70 percent. Note that
the second, weaker, component of H\,{\sc i} 
Ly\,$\alpha$ and Ly\,$\beta$ absorption associated
with the high-ion absorption is blended, so that the true component structure
of the H\,{\sc i} and the relative H\,{\sc i} column densities and H\,{\sc i}
$b$-values remain somewhat uncertain. The blending aspect
is not taken into account in the formal error estimate for $N$ and $b$ given in
Table 1, which is based on the profile fitting. While for 
the stronger of these components the absorption of O\,{\sc vi}, C\,{\sc iv}, 
and H\,{\sc i} is well aligned, there appears to be a small ($< 10$ km\,s$^{-1}$) 
velocity shift between H\,{\sc i} and the high ions in the
weaker component (see Table 1). If real, this shift may indicate that the H\,{\sc i}
and the metal ions may not trace the same gas phase in the weaker absorption 
component. Because of the blending of the H\,{\sc i} absorption, however, 
the reality of this shift remains unclear.

We derive for the column densities, listed in Table 1, log $N$(O\,{\sc vi}$)\approx 14.3$,
log $N$(C\,{\sc iv}$)\approx 13.1$, and log $N$(H\,{\sc i}$)\approx 13.4$ 
in the stronger of the two components. The resulting ion-to-hydrogen
ratios of $N$(O\,{\sc vi}$)/N$(H\,{\sc i}$)\sim 8$ and 
$N$(C\,{\sc iv})/$N$(H\,{\sc i}$)\sim 0.5$ already indicate 
that the metallicity of this absorber must be fairly high
\citep{Bergeron05}. Note that because of the 
blending problem in the Ly\,$\alpha$ and Ly\,$\beta$ lines 
the H\,{\sc i} column density may be regarded as upper limit,
so that the ratios given above could be even higher.


\begin{table}[t!]
\caption{H\,{\sc i} Ly\,$\alpha$ fit parameters for the absorption system at $z=2.1660$}
\begin{small}
\begin{tabular}{ccccc}
\hline
\hline
 & $z$ & log\,[$N$(H\,{\sc i}) (cm$^{-2}$)] & $b$ [km\,s$^{-1}$]  \\
\hline
\multicolumn{4}{c}{\it high resolution data} \\ \hline
1 &  2.16466 & 13.27 ($\pm$0.01) & 40.0 ($\pm$1.2) \\
2 &  2.16552 & 14.49 ($\pm$0.07) & 29.9 ($\pm$1.2) \\
3 &  2.16605 & 15.18 ($\pm$0.03) & 34.9 ($\pm$0.5) \\
4 &  2.16746 & 13.16 ($\pm$0.03) & 19.2 ($\pm$0.6) \\
5 &  2.16765 & 13.77 ($\pm$0.01) & 45.6 ($\pm$0.3) \\
\hline
\multicolumn{4}{c}{\it intermediate resolution data}   \\ \hline
1 & 2.16467  & 13.21 ($\pm$0.02) & 37.4 ($\pm$1.3) \\
2 & 2.16559  & 14.57 ($\pm$0.12) & 31.9 ($\pm$1.7) \\
3 & 2.16607  & 15.13 ($\pm$0.05) & 35.5 ($\pm$0.8) \\
4 & 2.16749  & 13.19 ($\pm$0.02) & 18.8 ($\pm$0.5) \\
5 & 2.16768  & 13.76 ($\pm$0.01) & 46.8 ($\pm$0.3) \\
\hline
\end{tabular}
\end{small}
\end{table}


\begin{table*}[t!]
\caption{Fit parameters for metal absorption in the absorber at $z=2.1660$}
\begin{tiny}
\begin{tabular}{ccccllllll}
\hline
\hline
\\
& \multicolumn{3}{c}{\small $z$} & \multicolumn{2}{c}{\small O\,{\sc vi}} &
\multicolumn{2}{c}{\small C\,{\sc iv}}&
\multicolumn{2}{c}{\small C\,{\sc iii}} \\
& O\,{\sc vi} & C\,{\sc iv} & C\,{\sc iii} & log[$N$(cm$^{-2}$)]
& $b$\,[km\,s$^{-1}$] & log[$N$(cm$^{-2}$)] & $b$\,[km\,s$^{-1}$]
& log[$N$(cm$^{-2}$)] & $b$\,[km\,s$^{-1}$] \\
\hline
\multicolumn{10}{c}{\it high resolution data} \\ \hline
1 & 2.16518 & ---     & ---     & 13.35($\pm$0.03)     & 15.4($\pm$1.3)
& ---         & ---            & ---                   & ---           \\
2 & 2.16542 & ---     & ---     & 13.02($\pm$0.08)     &  8.0($\pm$1.2)
& ---         & ---            & ---                   & ---           \\
3 & 2.16569 & 2.16569 & 2.16566 & 13.63($\pm$0.02)     & 13.5($\pm$0.7)
& 12.18($\pm$0.05) & 11.1($\pm$1.7)      & 12.68($\pm$0.07)           & 11.1$^{\rm a}$ \\
4 & 2.16600 & 2.16600 & 2.16602 & 13.41($\pm$0.03)     &  9.9($\pm$0.6) & 13.17($\pm$0.05)
&  9.3($\pm$0.5)      & 13.49($\pm$0.05)       & 15.1($\pm$1.5)     \\
5 & ---     & 2.16616 & 2.16618 & ---             & ---
& 12.96($\pm$0.10) & 10.0($\pm$1.6)      & 12.78($\pm$0.23)           &  8.0($\pm$3.7)     \\
6 & 2.16638 & 2.16638 & 2.16633 & 13.21($\pm$0.04)     & 13.7($\pm$0.8)
& 12.53($\pm$0.09) & 12.6($\pm$2.2)      & 12.22($\pm$0.47)       & 12.6$^{\rm a}$\\
7 & 2.16744 & 2.16741 & 2.16740 & 12.75($\pm$0.15) &  6.8($\pm$0.8)
& 11.90($\pm$0.05) &  4.8$^{\rm b}$ & 12.17($\pm$0.14) &  4.8$^{\rm a}$\\
8 & 2.16766 & 2.16770 & 2.16780 & 13.01($\pm$0.16) & 10.6($\pm$0.4)
& 11.84($\pm$0.08) &  9.2$^{\rm c}$ & 12.32($\pm$0.15) &  9.2$^{\rm a}$\\
9 & 2.16796 & 2.16792 & 2.16789 & 13.30($\pm$0.08) & 10.9($\pm$1.3)
& 11.50($\pm$0.12) &  4.8$^{\rm b}$ & 11.85($\pm$0.36) &  4.8$^{\rm a}$\\
\hline
\multicolumn{10}{c}{\it intermediate resolution data} \\ \hline
1 & 2.16521 & ---     & ---     & 13.12($\pm$0.03)     &  8.3($\pm$1.0)
& ---         &  ---           & ---                   & ---           \\
2 & 2.16544 & ---     & ---     & 13.13($\pm$0.04)     &  6.4($\pm$0.9)
& ---         &  ---           & ---                   &  ---          \\
3 & 2.16600 & 2.16567 & 2.16557 & 13.53($\pm$0.05)     & 10.6($\pm$0.9)
& 12.16($\pm$0.05) & 11.1($\pm$1.5)      & 12.87($\pm$0.11)           & 14.4($\pm$3.7)     \\
4 & 2.16606 & 2.16602 & 2.16603 & 13.73($\pm$0.04)     & 24.8($\pm$3.1)
& 13.23($\pm$0.02) & 10.4($\pm$0.4)      & 13.39($\pm$0.08)           & 10.4$^{\rm a}$\\
5 & ---     & 2.16619 & 2.16626 & ---             & ---
& 12.73($\pm$0.09) &  7.5($\pm$0.9)      & 13.12($\pm$0.15)           &  7.5$^{\rm a}$\\
6 & 2.16646 & 2.16637 & 2.16635 & 12.97($\pm$0.11)     & 10.4($\pm$2.0)
& 12.67($\pm$0.08) & 15.3($\pm$2.6)      & ---                   &  ---   \\
7 & 2.16748 & 2.16742 & ---     & 12.98($\pm$0.03)     &  7.3($\pm$0.9)
& 11.67($\pm$0.17) &  4.8($\pm$2.3)      & ---                   &  ---   \\
8 & 2.16767 & 2.16758 & ---     & 12.93($\pm$0.05)     &  2.8($\pm$0.9)
& 12.24($\pm$0.07) & 22.2($\pm$3.1)      & ---                   &  ---   \\
9 & 2.16796 & 2.16797 & ---     & 13.50($\pm$0.01)     & 12.7($\pm$0.6)
& 11.85($\pm$0.06) &  4.8($\pm$1.2)      & ---                   &  ---   \\
\hline
\end{tabular}
\end{tiny}
\tiny$^{\rm a}$ Fixed to $b_{\rm C\,{III}} = b_{\rm C\,{IV}}$ \\
\tiny$^{\rm b}$ Fixed to $b$-value derived from the intermediate resolution data \\
\tiny$^{\rm c}$ Lower limit, fixed to the minimal value \\
\end{table*}


\subsection{The O\,{\sc vi} system at $z=2.1660$}

The O\,{\sc vi} system at $z=2.1660$ exhibits a significantly more
complex absorption pattern than the absorber at $z=2.1098$, as can
be seen in the velocity profiles presented in Fig\,2. 
O\,{\sc vi} absorption is observed in eight individual absorption
components, spanning a velocity range as large as $\sim 300$ 
km\,s$^{-1}$. From the visual inspection it is further evident that
the absorption pattern of O\,{\sc vi} is different than those of 
the other detected intermediate and high ions (C\,{\sc iii}, C\,{\sc iv})
and H\,{\sc i}, although some of the components appear to be 
aligned in velocity space. As for the system at $z=2.1098$, there are no
significant differences in the absorption characteristics 
between the high-resolution data and the intermediate-resolution data.
However, the S/N ratio is somewhat lower in the latter for the
lines that are located in the blue part of the spectrum,
so that the resulting fitting values for $N$, $b$, and $z$
for the individual absorption components differ slightly
(Tables 2 and 3). 

We have modeled the H\,{\sc i} absorption by 
simultaneously fitting Ly\,$\alpha$ and Ly\,$\beta$ in four absorption 
components (components $2-5$; see Table 2), obtaining column densities 
between $13.2 <$ \,log $N$(H\,{\sc i})$< 15.2$.
One additional component (component 1) is present in the Ly\,$\alpha$ 
absorption, but is blended in Ly\,$\beta$ (see Fig.\,2), so that $N$(H\,{\sc i})
was derived solely from Ly\,$\alpha$. 
Note that for the H\,{\sc i} fit we have not tried to tie the
H\,{\sc i} component structure to the structure seen in the 
the metal ions, as this requires knowledge about the physical
conditions in the absorber. This aspect will be discussed in
detail in Sect.\,4.2, where we try to reconstruct the H\,{\sc i}
absorption pattern based on a photoionization model. Instead,
we have fitted the H\,{\sc i} absorption with the minimum 
number of absorption components required to match the observations
(Fig.\,2, lowest panel) and to obtain an estimate on the
total H\,{\sc i} column in the absorber.

By summing over the column densities in the 
individual absorption components we derive total column 
densities of log $N$(O\,{\sc vi}$)\approx 14.2$, 
log $N$(C\,{\sc iii}$)\approx 13.7$, log $N$(C\,{\sc iv}$)\approx 13.5$,
and log $N$(H\,{\sc i}$)\approx 15.3$. The resulting 
ion-to-hydrogen ratios of 
$N$(O\,{\sc vi}$)/N$(H\,{\sc i}$)\sim 0.1$ and
$N$(C\,{\sc iv}$)/N$(H\,{\sc i}$)\sim 0.02$ 
(representing the average over all components) 
are substantially smaller than in the $z=2.1098$ system,
pointing toward a lower (mean) metallicity of the absorber.

The complexity in the absorption pattern of the 
various species in this system and the large velocity spread 
suggests that this absorber arises in an extended multi-phase
gas structure.


\begin{table*}[]
\caption{Modelled column densities for the absorber at $z = 2.1098$}
\begin{small}
\begin{tabular}{crcccccccc}
\hline
\hline
& & \multicolumn{3}{c}{log [$N$\,(cm$^{-2}$)]} \\
& $v$\,[km\,s$^{-1}$] & C\,{\sc iv} & O\,{\sc vi} &
H\,{\sc i} &  log [$n_{\rm H}\,$(cm$^{-3}$)] & log $Z$
& log [$T$(K)] & $L$ [kpc] & $f_{\rm HI}$ \\
\hline
1 &     0 & 13.12 & 14.27 & 13.38           & $-$4.20
& $-$0.24 & 4.54 & 19.9 & $-$5.21 \\
2 & $+$25 & 12.23 & 13.50 & 13.37           & $-$4.25
& $-$1.02 & 4.64 & 30.5 & $-$5.35 \\
\hline
2 & $+$25 & 12.23 & 13.50 & 12.57$^{\rm a}$ & $-$4.28
& $-$0.24 & 4.57 &  4.7 & $-$5.32 \\
\hline
\end{tabular}
\end{small}
\\
\tiny$^{\rm a}$ Our best H\,{\sc i} guess in the model for the second
component with fixed metallicity\\
\end{table*}


\section{Ionization modeling and physical conditions in the gas}

To infer information on the physical properties of the
two O\,{\sc vi} absorbers towards PKS\,1448$-$232 we have modeled 
in detail the ionization conditions in these systems. Since
the two absorbers at $z=2.1098$ and $z=2.1660$ have redshifts
close to the quasar redshift ($z_{\rm QSO}=2.208$), it is
necessary to check whether the two systems lie in the 
proximity zone of the background quasar and are influenced
by its ionizing radiation.

With the above given redshifts, the two absorbers have 
velocity separations from the QSO of $\delta v_{2.1098}\approx 9000$ km\,s$^{-1}$ 
and $\delta v_{2.1660}\approx 4000$ km\,s$^{-1}$ and thus 
the absorber at $z=2.1660$ can be regarded (depending on the definition)
as an associated system.
With a (monochromatic) luminosity at the Lyman Limit of
$L_{912}=3.39 \times 10^{31}$ erg\,s$^{-1}$\,Hz$^{-1}$, the
size of the sphere-of-influence of the ionizing radiation from
PKS\,1448$-$232 is known to be $6.7$ Mpc, corresponding to a velocity
separation of $\sim 1400$ km\,s$^{-1}$ \citep{Fox08}.  
Therefore, it is safe to assume that the ionizing radiation coming
from PKS\,1448$-$232 itself has no measurable influence on the 
ionization conditions in the two O\,{\sc vi} systems.

The small $b$-values measured for O\,{\sc vi}, C\,{\sc iv}, 
and H\,{\sc i} indicate that collisional
ionization is not the origin for the presence of O\,{\sc vi}
in the gas. It is common to assume that the observed Doppler parameters
($b_{\rm obs}$) are composed of a thermal and a turbulent component
($b_{\rm th}$ and $b_{\rm turb}$, respectively), so that
$b^{2}_{\rm obs} = b^{2}_{\rm th} + b^{2}_{\rm turb}$. The thermal
component can be expressed by $b^{2}_{\rm th} = 2kT/m$, where
$T$ is the gas temperature and $m$ is the mass of the considered ion. 
The Doppler parameters measured for the O\,{\sc vi} components 
in the two absorbers are all $b<16$ km\,s$^{-1}$ and many
of them are $b<10$ km\,s$^{-1}$ (see Tables 1 and 3), indicating 
that $T<10^5$ K. This value is below the peak temperature
of O\,{\sc vi} in a collisional ionization equilibrium 
 \citep[$T\sim 3\times 10^5$ K;][]{Sutherland93}; 
it is also lower than the temperature range expected for
O\,{\sc vi} arising in turbulent mixing layers in the
interface regions between cold and hot gas 
\citep[$T=10^5-10^6$ K;][]{Kwak10}. Consequently, photoionization by the 
hard UV background remains as the only plausible origin for the 
presence of O\,{\sc vi} in the two high-ion absorbers towards
PKS\,1448$-$232.

Based on these considerations, we have modeled the ion column densities
in the two O\,{\sc vi} systems using the photoionization code CLOUDY 
\citep[version C08;][]{Ferland}. For this, we have assumed a solar relative
abundance pattern of O and C and an optically thin plane-parallel geometry 
in photoionization equilibrium, 
exposed to a \citet{HM01} UV background spectrum at $z = 2.16$, which is
normalized to $\log~J_{912} = -21.15$ \citep{Scott} at the Lyman limit.
 
We further assume that each of the observed velocity components is 
produced by a "cloud", which we model as an individual entity.
As input parameters we consider the measured column densities 
of C\,{\sc iii} (only for the $z=2.1660$ absorber), C\,{\sc iv},  
O\,{\sc vi}, the metallicity $Z$ (in solar units), 
and the hydrogen particle density $n_{\rm H}$. The metallicity 
of each cloud and the hydrogen density were varied in a range 
appropriate for intergalactic clouds  (i.e., $-3\leq \,$log$~Z\,\leq 0$
and $-5\leq \,$log$ ~n_{\rm H} \leq 0$).  

We then applied the following iterative modeling procedure. 
In a first step, CLOUDY was run with a set of values of $Z$, $n_{\rm H}$ and 
$N$(H\,{\sc i}), where $N$(H\,{\sc i}) is constrained by the 
observations. In a second step, the corresponding
values of $N$(C\,{\sc iii}), $N$(C\,{\sc iv}), and $N$(O\,{\sc vi})
were calculated. The output was compared with the observed 
column densities and, in case of a mismatch, the input parameters
$Z$ and $n_{H}$ were adjusted for the next iteration step. 
This process was repeated until the differences between the 
output column densities and the observed values became negligible
and we obtained a unique solution.
Next to the ion column densities, our CLOUDY model provides 
information on the neutral hydrogen fraction, $f_{\rm HI}$, 
the gas temperature, $T$, and the absorption pathlength,
$L=N$(H\,{\sc i}$)/(f_{\rm HI}\,n_{\rm H})$. 

\subsection{The system at $z=2.1098$}

As mention earlier, absorption by O\,{\sc vi} and C\,{\sc iv} is 
well aligned in both components in this system, while the true
component structure of the H\,{\sc i} is uncertain because of 
blending effects in the Ly\,$\alpha$ and Ly\,$\beta$ lines.
Because of the alignment of O\,{\sc vi} and C\,{\sc iv} 
we assume a single-phase model, in which each of the two components
(clouds) at $v=0$ and $+25$ km\,s$^{-1}$ in the $z=2.1098$ rest 
frame hosts O\,{\sc vi}, C\,{\sc iv}, and H\,{\sc i} at
column densities similar to the ones derived from the profile
fitting. Consequently, we have chosen log $N$(H\,{\sc i}$)=13.37$ and $13.38$
as input for the CLOUDY modeling and followed the procedure
outlined above. The results from the CLOUDY modeling
of the $z=2.1098$ absorber are summarized in Table\,4. Our model
reproduces well the observed O\,{\sc vi} and C\,{\sc iv} column
densities in both components, if the clouds have a density of log $n_{\rm H}
\approx -4.2$, a temperature of log $T\approx 4.6$, and a
neutral hydrogen fraction of log $f_{\rm HI}\approx -5.3$. However,
to match the observations, the second component (at $+25$ km\,s$^{-1}$)
in our initial model (Table 4, first two rows) 
needs to have metallicity of log $Z=-1.02$, which is
$\sim 0.8$ dex lower than for the other component (log $Z=-0.24$). 
The absorption path lengths are $\sim 20$ kpc for the component
at $0$ km\,s$^{-1}$ and $\sim 30$ kpc for the component
at $+25$ km\,s$^{-1}$. 

Because of the blending problem in the H\,{\sc i} Ly\,$\alpha$ and Ly\,$\beta$
absorption, which affects particularly the estimate for $N$(H\,{\sc i})
in the cloud at $+25$ km\,s$^{-1}$ (Fig.\,1), we have set up a second
CLOUDY model in which we have tied the metallicity of the $+25$ km\,s$^{-1}$
component to the metallicity of the other component (log $Z=-0.24$),
but leaving $N$(H\,{\sc i}) for this component as a free parameter. 
>From this we derive a value of log $N$(H\,{\sc i}$)=12.57$ for the cloud
at $+25$ km\,s$^{-1}$ and the absorption path-length reduces to
$L=4.7$ kpc. In view of the blending, we regard this model
as more plausible compared to the model with two different metallicities
and a larger absorption path-length.

Summarizing, our CLOUDY modeling suggests that the $z=2.1098$ absorber towards
PKS\,1448$-$232 represents at relatively simple, metal-rich O\,{\sc vi}
absorber in which the high ions O\,{\sc vi} and C\,{\sc iv} coexist in
a single gas-phase.


\begin{table*}[t!]
\caption{Modelled column densities for the C\,{\sc iii}/C\,{\sc iv} absorbing 
phase in the $z=2.1660$ absorber}
\begin{scriptsize}
\begin{tabular}{clccccccccccccccc}
\hline
\hline
& & \multicolumn{4}{c}{log [$N$\,(cm$^{-2}$)]} \\
& $v$\,[km\,s$^{-1}$] & C\,{\sc iii} & C\,{\sc iv} & O\,{\sc vi} &
H\,{\sc i}$^{\rm a}$ & log [$n_{\rm H}\,$(cm$^{-3}$)] & log $Z$ & log [$T$(K)] & $L$ [kpc] & $f_{\rm H}$ \\
\hline
3 & $-$28  & 12.68 & 12.18 & 10.97          & 14.51 & $-$2.74 & $-$1.7 & 4.42 & 0.3   & $-$3.68  \\
4 & $+$0   & 13.49 & 13.17 & 12.25          & 14.18 & $-$2.97 & $-$1.7 & 4.46 & 4.1   & $-$3.95  \\
5 & $+$16  & 12.78 & 12.96 &  ---           & 14.51 & $-$3.56 & $-$1.7 & 4.58 & 16.3  & $-$4.63  \\
6 & $+$37  & 12.22 & 12.53 & 12.86          & 14.08 & $-$3.71 & $-$1.7 & 4.61 & 12.3  & $-$4.79  \\
7 & $+$134 & 12.17 & 11.90 & 10.92          & 13.26 & $-$2.93 & $-$1.0 & 4.38 & 0.04  & $-$3.84  \\
8 & $+$162 & 12.32 & 11.84 & 10.49          & 13.57 & $-$2.67 & $-$1.0 & 4.34 & 0.02  & $-$3.55  \\
9 & $+$182 & 11.85 & 11.50 & 10.38          & 12.99 & $-$2.84 & $-$1.0 & 4.37 & 0.01  & $-$3.73  \\
\hline
\end{tabular}
\end{scriptsize}
\\
\tiny$^{\rm a}$ Our best H\,{\sc i} guess in the models.\\
\end{table*}


\subsection{The system at $z=2.1660$}

We started to model this system with CLOUDY, again under the assumption of a 
single gas-phase hosting the observed intermediate and high ions 
C\,{\sc iii}, C\,{\sc iv}, and O\,{\sc vi} in the various
subcomponents. However, during the modeling process 
it quickly turned out that it is impossible to match the observed 
column densities of C\,{\sc iii} and O\,{\sc vi} in a single gas-phase
in the components, where these two ions are aligned
in velocity space. Our modeling instead indicates that 
the C\,{\sc iii} absorption must arise in an
environment that has a relatively high gas density and that is 
spatially distinct from the O\,{\sc vi} phase. In a second
step, we have tried to tie together the high ions C\,{\sc iv} and
O\,{\sc vi} in a single gas phase (as for the $z=2.1098$ system) 
in the relevant absorption components, ignoring the C\,{\sc iii} phase. 
Again, this approach does not deliver satisfying results, as
we obtain for some of the components, for which C\,{\sc iv}/O\,{\sc vi} 
is constrained by observations, very low gas densities and
very large absorption pathlengths on Mpc scales, which
are highly unrealistic. Given the fact, that the overall
component structure of O\,{\sc vi} and C\,{\sc iv} 
are substantially different in this system (Fig.\,2), this
result is not really surprising.

The only modeling approach, for which we obtain realistic results on
gas densities, temperatures and absorption path-lengths in this
system and its subcomponents is a two-phase model, in which C\,{\sc iii}
coexists with C\,{\sc iv} and part of the H\,{\sc i} in one (spatially 
relatively confined) phase, and O\,{\sc vi} and part of the H\,{\sc i} 
in a second (spatially relatively extended) phase. The coexistence
of C\,{\sc iii} and C\,{\sc iv} in one phase is further suggested
by the fact that C\,{\sc iii} and C\,{\sc iv} absorption is well 
aligned in velocity space (see Fig.\,2). The results from this 
two-phase model are presented in Tables 5 and 6. A critical issue
for the modeling of this complex multi-phase absorber with its
many absorption components is an assumption for the neutral
gas column density in each subcomponent (and phase). Since in the
H\,{\sc i} Ly\,$\alpha$ and Ly\,$\beta$ absorption most subcomponents 
are smeared together to one large absorption trough, the 
observational data provide little information on the distribution
of the H\,{\sc i} column densities among the individual components.
Yet, the data give a solid estimate for the {\it total} H\,{\sc i} column
density in the absorber (log $N\approx 15.3$; see Sect.\,2.4), which 
must match the sum of $N$(H\,{\sc i}) over all subcomponents considered
in our model. Consequently, we included in our iteration procedure
the constraints on $N$(H\,{\sc i}$)_{\rm tot}$ and the {\it shape}
of the (total) H\,{\sc i} absorption profile. 
The latter aspect also concerns the choice of the gas temperature
in the model, as $T$ regulates the thermal Doppler-broadening and thus
the width of the modeled H\,{\sc i} lines. We have modeled the H\,{\sc i}
width following the aproach of \citet{Ding}.

With these various constraints we first modeled the C\,{\sc iii}/C\,{\sc iv} phase
in the absorber. However, because of the extremely complex parameter
space, we did not find a unique solution for $(T,n_{\rm H},Z)$ among the 
individual components, but had to make further constraints. Since
the individual components observed in C\,{\sc iii}/C\,{\sc iv} 
are very close together in velocity space, we assumed they all have
the same metallicity and, based on the $Z$ range allowed in the
model, we set log $Z=-1.5$ for all subcomponents.
This model was able to match the observed column densities
of these two ions in the individual subcomponents, but did not match 
well the gross shape of the overall H\,{\sc i} absorption, suggesting
that the metallicity in this absorber is non-uniform among the
individual absorption components.
Therefore, we refined our model, now using two different 
metallicities: log $Z=-1.7$ for the saturated H\,{\sc i} components
and log $Z=-1.0$ for the weaker H\,{\sc i} components (see Tables 5 and 6
for details). Although not perfect, this model delivers a satisfying
match between the modeled spectrum and the UVES data.

Adopting this model, we find that the C\,{\sc iii}/C\,{\sc iv} absorbing
components have temperatures between
log $T=4.3$ and $4.6$, densities between log $n_{\rm H}=-3.7$
and $-2.7$, and neutral gas fractions between log $f_{\rm HI}=-4.8$
and $-3.6$ (see Table 5).
The absorption path lengths vary between $0.3$ and $16.3$ kpc 
for the components with log $Z=-1.7$, and between $0.01$ and 
$0.04$ kpc for the components with log $Z=-1.0$. These numbers 
suggest that the C\,{\sc iii}/C\,{\sc iv}
absorbing phase resides in relatively small and confined 
gas clumps. This scenario is supported by the small turbulent
$b$-values of $<6$ km\,s$^{-1}$ for the subcomponents 
that we derive in our model. Note that in Table 5 we also
list the predicted column densities for O\,{\sc vi}, 
which are typically $1-2$ orders of magnitude below 
the observed ones in this absorber. This, again, underlines
that C\,{\sc iii}/C\,{\sc iv} and O\,{\sc vi} must reside
in different gas phases with different physical conditions 
to explain the observed column densities. 

Finally, we have modeled the O\,{\sc vi} absorbing phase in the
$z=2.1660$ absorber, based on the observed O\,{\sc vi} 
column densities. Since there are no ions other 
than H\,{\sc i} and O\,{\sc vi} that could provide information
about the physical conditions in this phase, we fixed
the metallicity of the gas to log $Z=-1.7$ and 
log $Z=-1.0$ (equal to the C\,{\sc iii}/C\,{\sc iv} phase) 
and constrained the temperature range [$T_{\rm min},T_{\rm max}$] 
in the CLOUDY models based on the observed line widths 
of O\,{\sc vi} (giving $T_{\rm max}$) and the 
modeling results of the C\,{\sc iii}/C\,{\sc iv} phase
(giving $T_{\rm min}$ for all components except the 
first two). The results of this model are shown in Table 6. 
We derive gas densities in the range log $n_{\rm H}=-4.6$ to 
$-3.2$ and neutral gas fractions in the range 
log $f_{\rm HI}=-5.8$ to $-4.6$. The absorption path length varies 
between $19.8$ and $83.3$ kpc for the components with log $Z=-1.7$, 
and between $1.3$ and $38.3$ kpc for the ones with log $Z=-1.0$.
The mismatch in $N$(O\,{\sc vi}) between the model and the 
data for components one and nine (see Table 6) points towards
a metallicity distribution among the individual
absorption components that is even more complex than the one 
assumed in our model. Despite this (minor) concern, 
our CLOUDY modeling for O\,{\sc vi} provides clear
evidence that the O\,{\sc vi} absorbing phase has substantially lower
gas densities than the C\,{\sc iii}/C\,{\sc iv} absorbing
phase and is spatially more extended.

In summary, our CLOUDY modeling of the $z=2.1660$ absorber
suggests that this system represents a complex
multi-phase gas structure, in which a number of cooler,
C\,{\sc iii}/C\,{\sc iv} absorbing cloudlets are embedded
in a spatially more extended, O\,{\sc vi} absorbing gas phase
spanning a total velocity range of $\sim 300$ km\,s$^{-1}$.
Although the metallicity is not well constrained in our
model, it suggests that log $Z\leq -1$ in the absorber, which
is $\sim 0.8$ dex below the value obtained for the
system at $z=2.1098$.


\begin{table*}[t!] 
\caption{Modelled column densities for the O\,{\sc vi} absorbing
phase in the $z=2.1660$ absorber}
\begin{scriptsize}
\begin{tabular}{clcccccrc} 
\hline
\hline 
& & \multicolumn{2}{c}{log [$N$\,(cm$^{-2}$)]} \\
& $v$\,[km\,s$^{-1}$] & O\,{\sc vi} &
H\,{\sc i}$^{\rm a}$ &  log [$n_{\rm H}\,$(cm${-3}$)] & log $Z$ & log [$T$(K)] & $L$ [kpc] & $f_{\rm H}$ \\
\hline
\hline
1 & $-$78    & 12.96$^{\rm b}$ & 13.17 & $< -$3.76 & $-$1.7 & $<$5.36  & $<$19.8  &  $>-$5.86  \\
2 & $-$55    & 13.02       & 14.10 & $< -$3.77 & $-$1.7 & $<$4.79  & $<$25.8  &  $>-$5.03  \\  
3 & $-$30    & 13.63       & 14.51 & $-$3.91 ... $-$3.31 & $-$1.7 & 4.42 ... 5.24  &58.4  ... 67.7  & $-$5.50 ... $-4.84$  \\   
4 & $+$0     & 13.41       & 14.18 & $-$3.99 ... $-$3.88 & $-$1.7 & 4.46 ... 4.98  &42.6  ... 83.3  & $-$5.36 ... $-$4.95  \\ 
6 & $+$36    & 13.21       & 14.08 & $-$3.92 ... $-$3.23 & $-$1.7 & 4.61 ... 5.26  &32.1  ... 21.5  & $-$5.52 ... $-$5.00  \\
7 & $+$136   & 12.75       & 13.26 & $-$3.72 ... $-$3.72 & $-$1.0 & 4.38 ... 4.64  & 1.3  ... 2.0   & $-$4.82 ... $-$4.61  \\ 
8 & $+$157   & 13.01       & 13.57 & $-$3.70 ... $-$3.52 & $-$1.0 & 4.34 ... 5.03  & 2.2  ... 6.1   & $-$5.19 ... $-$4.56  \\ 
9 & $+$185   & 13.25$^{\rm c}$ & 12.99 & $-$4.60 ... $-$4.25 & $-$1.0 & 4.37 ... 5.06  & 32.3  ... 38.3  & $-$5.76 ... $-$5.48  \\ 
\hline  
\end{tabular} 
\end{scriptsize}
\\
\tiny$^{\rm a}$ Our best H\,{\sc i} guess in the models\\ 
\tiny$^{\rm b}$ Observed {log} $N$(O\,{\sc vi}) $= 13.35$\\ 
\tiny$^{\rm c}$ Observed {log} $N$(O\,{\sc vi}) $= 13.30$\\  
\end{table*}  


\section{Discussion}

Our detailed analysis of the two O\,{\sc vi} absorbers 
at $z=2.1098$ and $z=2.1660$ towards the quasar 
PKS\,1448$-$232 displays the large diversity and complexity of 
high-ion absorbers at high redshift. 

During the past years, a number of studies using both
optical observations \citep[e.g.,][]{Bergeron02, Carswell02, 
Simcoe02, Simcoe04, Simcoe06, Bergeron05, Aguirre08} 
and numerical simulations \citep[e.g.,][]{Fangano07, 
Kawata07} have been dedicated to investigate
the properties of high-redshift O\,{\sc vi} systems 
and their relation to galaxies.

Based on their survey of O\,{\sc vi} absorbers in the 
redshift range $z=2.0-2.6$, \citet{Bergeron05}
suggested that O\,{\sc vi} systems may be classified into
two different populations: metal-rich absorbers (``type 1'')
that have large $N$(O\,{\sc vi})/$N$(H\,{\sc i}) ratios 
and that appear to be linked to galaxies and 
galactic winds, and metal-poor absorbers (``type 0'')
with small $N$(O\,{\sc vi})/$N$(H\,{\sc i}) ratios that
trace the intergalactic medium.
The two absorbers towards PKS\,1448$-$232 discussed in this paper
do not match the classification scheme of \citet{Bergeron05}. The absorber
at $z=2.1098$ has a very large $N$(O\,{\sc vi})/$N$(H\,{\sc i})
ratio of $\sim 8$ (i.e., it is of type 1); it is a simple, single-phase,
metal-rich system with a metallicity slightly below the solar value.
Yet, this system is completely isolated with no strong H\,{\sc i} Ly\,$\alpha$
absorption within $1000$ km\,s$^{-1}$. In contrast, the absorber
at $z=2.1660$ has a $N$(O\,{\sc vi})/$N$(H\,{\sc i}) ratio
of only $\sim 0.1$ and a metallicity of $0.1$ solar or lower
\citep[i.e., it is of type 0 according to][]{Bergeron05}.
However, this absorber is a complex multi-phase system
with a non-uniform metallicity, suggesting that originates
in a circumgalactic environment.  
While this mismatch with the \citet{Bergeron05} classification
scheme certainly has no statistical relevance for the general
interpretation of O\,{\sc vi} absorbers at high redshift,
the results suggest that for a thorough understanding
of highly-ionized gas at high redshift
the absorption characteristics
of O\,{\sc vi} systems may be too diverse for
a simple classification scheme based solely on observed
(and partly averaged) column density ratios of O\,{\sc vi},
H\,{\sc i} and other ions. 

One critical drawback of many
previous O\,{\sc vi} surveys at high $z$ is that they 
often consider only simplified models for the ionization
conditions in their sample of high-ion absorbers, so that 
the multi-phase character of the gas and possible ionization 
conditions far from a photoionization equilibrium 
are only insufficiently taken into account.
As pointed out by \citet{Fox11}, single-phase, 
single-component ionization models, if applied, 
will deliver physically irrelevant results for most
of the O\,{\sc vi} systems at high $z$.
This implies that previous estimates of the
baryon- and metal-content of O\,{\sc vi} absorbers at
low and high $z$ possibly are afflicted with large systematic
uncertainties.

One firm conclusion from many previous observational and theoretical
studies of high-ions absorbers is that a considerable 
fraction of the O\,{\sc vi} systems at low and high $z$ 
must arise in the metal-enriched circumgalactic
environment of (star-forming) galaxies \citep[e.g.,][]{Wakker09, 
Prochaska11, Fox11a, Tepper-Garcia11, Fangano07}. Thus, the complex
absorption pattern observed in the $z=2.1660$ system 
towards PKS\,1448$-$232 and many other O\,{\sc vi} absorbers
at high $z$ may reflect the complex gas distribution of enriched
gaseous material that was ejected from galaxies into the IGM during
their wind-blowing phase \citep[e.g.,][]{Kawata07}.
In this context, \citet{Schaye07} suggested that the intergalactic 
metals have been transported from galaxies through galactic winds and reside
in the form of dense, small and high-metallicity patches within large
hydrogen clouds. These authors point out that much of the scatter in
the metallicities derived for high-redshift absorbers could be explained by
the spatially varying number of the metal-rich patches and the 
different absorption path lengths through the surrounding 
metal-poor intergalactic filament instead of an overall (large-scale)
metallicity scatter in the IGM. In this scenario, 
the substantial differences in the metallicities of the 
two O\,{\sc vi} systems towards PKS\,1448$-$232, and 
even the intrinsic metallicity variations within the
$z=2.1660$ system, 
could be explained by the different geometries of the absorbing structures,
suggesting that much of the H\,{\sc i} that is associated with the metal absorption 
in velocity space, arises in a spatially distinct region. A similar 
conclusion was drawn by \citet{Tepper-Garcia11}, who studied the 
nature of O\,{\sc vi} absorbers at low $z$ using a set of 
cosmological simulations. 
Note that also absorbers with larger H\,{\sc i} column densities,
such as Lyman-limit systems (LLS) and damped-Lyman $\alpha$
systems (DLAs), sometimes exhibit abundance variations among
the different velocity subcomponents \citep[e.g.,][]{Richter05,
Prochter10}. This indicates that the metals in the 
gas surrounding high $z$ galaxies are not well mixed.

The observed velocity differences between O\,{\sc vi} 
and other ions and the multi-phase nature of the gas provide
further evidence for an inhomogeneous metallicity and 
density distribution in intervening high-ion absorbers.
It is an interesting fact that 
the velocity misalignment appears to concern only
the O\,{\sc vi} absorbing phase in high-ion absorbers at high redshift,
while other high ions such as N\,{\sc v} and C\,{\sc iv} generally
appear to be well aligned with H\,{\sc i}, even in systems
that exhibit a complex velocity-component structure \citep{FR09}.
This puzzling aspect underlines
that additional detailed studies of individual O\,{\sc vi} absorption
systems could be very important for our understanding
of intergalactic and circumgalactic gas at
high redshift, as this ion traces a metal-enriched gas phase that 
cannot be observed by other means.

\section{Summary and outlook}

In this paper, we have investigated two O\,{\sc vi} absorbers at 
$z=2.1098$ and $z=2.1660$ towards the quasar PKS\,1448$-$232.
For this, we have used high- ($R\approx 75,000$) and 
intermediate-resolution ($R\approx 45,000$) optical spectra 
obtained with the VLT/UVES instrument and CLOUDY 
photoionization models. 

The O\,{\sc vi} system at $z=2.1098$ is characterized by 
strong O\,{\sc vi} absorption and weak H\,{\sc i} absorption 
in a relatively simple, two-component absorption pattern.
The absorption by O\,{\sc vi}, C\,{\sc iv}, and H\,{\sc i} are
well aligned in velocity space, indicating that they trace
the same gas phase. From a detailed photoionization
modeling of this system we derive a metallicity of $\sim 0.6$ 
solar, a characteristic density of log $n_{\rm H}
\approx -4.2$, a temperature of log $T\approx 4.6$,
and a total absorption path length of $\sim 30$ kpc. The absorber
is isolated with no strong H\,{\sc i} Ly\,$\alpha$ absorption
within $1000$ km\,s$^{-1}$.

The O\,{\sc vi} absorber at $z=2.1660$ represents a complicated,
multi-component absorption system with eight relatively weak and narrow
O\,{\sc vi} absorption components spanning almost 
$300$ km\,s$^{-1}$ in radial velocity. The O\,{\sc vi} 
components are accompanied by strong 
H\,{\sc i} absorption and C\,{\sc iii}, C\,{\sc iv} absorption. The 
O\,{\sc vi} component structure differs from that of H\,{\sc i}
and C\,{\sc iv}, indicating a multi-phase nature of the absorber.
Our photoionization modeling with CLOUDY suggests the presence of (at least) two 
distinct gas phases in this system. C\,{\sc iii}, C\,{\sc iv} and most 
of the H\,{\sc i} appear to coexist in several, relative
compact cloudlets at gas densities of 
log $n_{\rm H}\approx -3.7$ to $-2.7$, temperatures of 
log $T\approx 4.3-4.6$ and absorption path lengths of
$<16$ kpc. O\,{\sc vi} appears to reside in a
highly ionized, more extended gas phase at densities
in the range log $n_{\rm H}\approx -4.6$ to $-3.2$,
temperatures between log $T\approx 4.3$ and $5.3$,
and absorption path lengths up to $83$ kpc.
While the exact metallicity of the absorber is not well
constrained, our modeling favours a non-uniform
metal abundance among the individual absorption components
with (at least) two different metallicities of 
log $Z=-1.7$ and log $Z=-1.0$.

Our study displays the large diversity and complexity of
O\,{\sc vi} systems at high redshift. We speculate that 
some of the observed differences between the two high-ion
absorbers towards PKS\,1448$-$232 could be a result of 
a inhomogeneous metallicity and density distribution in 
the photoionized IGM. Our study indicates that
multi-phase, multi-component high-ion absorbers
like the one at $z=2.1660$ demand a detailed 
ionization modeling of the various subcomponents 
to obtain reliable information on physical conditions
and metal-abundances in the gas.
We conclude that a rather large effort is required to achieve 
a more complete view on the nature of O\,{\sc vi}
absorbers at high redshift. 

For the future, we are planning to 
continue our investigation on these systems by using a larger sample of 
O\,{\sc vi} absorbers in high-quality UVES archival
data and compare their absorption characteristics with 
artificial spectra generated from numerical simulations 
of star-forming galaxies and their intergalactic environment.

\begin{acknowledgements}

N.D. and P.R. acknowledge financial support by the German
\emph{Deut\-sche For\-schungs\-ge\-mein\-schaft}, DFG,
through grant Ri 1124/5-1. 

\end{acknowledgements}

\bibliographystyle{aa} 
\bibliography{papers.bib} 

\label{lastpage}

\end{document}